\documentclass[conference]{IEEEtran}
\usepackage{cite}
\usepackage{amsmath,amssymb,amsfonts}
\usepackage{algorithmic}
\usepackage{graphicx}
\usepackage{textcomp}
\usepackage{xcolor}
\def\BibTeX{{\rm B\kern-.05em{\sc i\kern-.025em b}\kern-.08em
    T\kern-.1667em\lower.7ex\hbox{E}\kern-.125emX}}

\usepackage{subcaption}
\usepackage{ctable}
\usepackage{bbold}

\graphicspath{{figures/}}

\newcommand*{\Eqref}[1]{Eq.~\eqref{#1}}
\def\*#1{\mathbf{#1}}
\begin{document}

\title{Dynamic Review-based Recommenders}


\author{\IEEEauthorblockN{Kostadin Cvejoski\IEEEauthorrefmark{1}\IEEEauthorrefmark{3}, Rams\'es J. S\'anchez\IEEEauthorrefmark{1}\IEEEauthorrefmark{2}, Christian Bauckhage\IEEEauthorrefmark{3} and C\'esar Ojeda\IEEEauthorrefmark{4}}
\IEEEauthorblockA{\IEEEauthorrefmark{1}Competence Center Machine Learning Rhine-Ruhr}
\IEEEauthorblockA{\IEEEauthorrefmark{2}B-IT, University of Bonn, Bonn, Germany}
\IEEEauthorblockA{\IEEEauthorrefmark{3}Fraunhofer Center for Machine Learning and Fraunhofer IAIS, 53757 Sankt Augustin, Germany}
\IEEEauthorblockA{\IEEEauthorrefmark{4}Berlin Center for Machine Learning and TU Berlin, 10587 Berlin, Germany\\
\{kostadin.cvejoski, christian.bauckhage\}@iais.fraunhofer.de,\\ojeda.marin@tu-berlin.de, sanchez@bit.uni-bonn.de}}

\maketitle

\begin{abstract}
Just as user preferences change with time, item reviews also reflect those same preference changes. In a nutshell, if one is to sequentially incorporate review content knowledge into recommender systems, one is naturally led to dynamical models of text. In the present work we leverage the known power of reviews to enhance rating predictions in a way that (i) respects the causality of review generation and (ii) includes, in a bidirectional fashion, the ability of ratings to inform language review models and vice-versa, language representations that help predict ratings end-to-end. Moreover, our representations are time-interval aware and thus yield a continuous-time representation of the dynamics. We provide experiments on real-world datasets and show that our methodology is able to outperform several state-of-the-art models.
Source code for all models can be found at \cite{source_code}.
\\
\end{abstract}


\section{Introduction}

Following the deep learning agenda, the success of modern recommender systems heavily relies on 
%
their ability
to leverage meaningful representations that allow for the accurate prediction of a purchase or a rating. Fundamentally one must dwell into  \textit{interest modeling}, as an effective recommendation is such that it uncovers, for a given user, a hidden interest in an unknown item. It is natural to
study 
the change of user interest with time and, in the present work, we seek to incorporate these dynamical notions with those of text reviews. 

Reviews are effectively a form of recommendation, and one that is directly provided by the user. The challenge however, stems from the unstructured and ambiguous nature of reviews (and natural language itself). A user might, for example, simultaneously highlight positive and negative aspects of the different items she reviews. Following current trends in natural language processing, we leverage review content through neural models of text
and attention mechanisms,
and guarantee the information content of those representations via reproduction quality.
We encourage the dynamical aspect of text recommendations by learning representations which help predict both \textit{when} is the next review arriving and \textit{what} does it say. 
One is then led naturally to dynamical language models, since enforcing good text predictions ensures its dynamical representation quality.

\section{Related Work}
There is a large body of research invested in recommender systems (RS), a big part of which has lately been devoted to 
capture the temporal dynamics of both users and items. 
One of the first temporal models for recommendation is the TimeSVD++ \cite{10.1145/1721654.1721677}, which extends the SVD++ matrix factorization algorithm by introducing time-dependent latent factors. 
From the neural network perspective, many models for RS have been developed
\cite{he2017neural, salakhutdinov2007restricted, sedhain2015autorec},
and Recurrent Neural Networks (RNNs) have been used to capture time-ordered user activity. For example, session-based item recommendation use RNNs to infers user preferences from sessions of user behaviour \cite{hidasi2015session, quadrana2017personalizing,tan2016improved,twardowski2016modelling}. 
Another example, closer to our work, is the Recurrent Recommender Networks (RRN) which uses two independent RNNs to model user and item dynamics separately \cite{RecurrentRecommender}.
%

Just as with user (and item) temporal representations, including review content representations has also been shown to significantly improve rating prediction and item recommendation \cite{catherine2017transnets, Dong2019AsymmetricalHN,seo2017interpretable,zheng2017joint}.
However, some of these models break \textit{causality}, in the sense that they either use the review of the item whose rate one is predicting, or use item reviews that have not been received by the time the item of interested was rated. 

Finally, a model that combines RRN (a dynamical RS) with character-based autoregressive language models for reviews has recently been develop \cite{wu2016joint}. This work however, does not leverage the review content for rating prediction.

In contrast to all these works, we combine dynamical recommender systems with a dynamical language model that captures review content evolution, and use the review representations together with the user-item temporal representations in a causal fashion, to predict the rating of the next review. 
%
%

\section{Dynamic Review-based Recommenders (DRR)}
\label{sec:Model}

\begin{figure*}[t!]
  \centering
  \includegraphics[width=.5\linewidth]{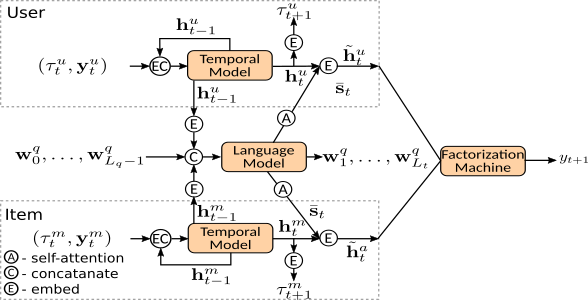}
  \caption{Dynamic Review-based Recommender. The model consists of three interacting components: (i) a temporal model composed of two RNNs, one for users and the other for items, which we called \textit{Dynamic Model of Review Sequences}; (ii) a neural language model which leverages the temporal representations of both user and items, and which we called \textit{Dynamic Model of Review Content}; and (iii) a \textit{Rating Model} which combines the user and item temporal representations with the review content representations to predict ratings. Note that when $q=t$ in the language model component, the Dynamic Review-based Recommender is causal. The model is non-causal when $q=t+1$.}
  \label{fig:model}
\end{figure*}


The interests and preferences of users 
vary as they age, or change their social status or lifestyle.
%
%
Exogenous factors like trends or seasons also affect user preferences.
For example, users tend to look for different clothe types in winter than those they look for in summer. Users also tend to change their music tastes as they age. 
Such preference changes are naturally encoded in the collections of reviews and ratings given by these users over time.
Our goal is to learn representations capturing them.
%
We therefore develop a model that explicitly uses the text content and ratings of past reviews together with the history of when those reviews were written to better predict user interest in unknown items. 
%

Consider a dataset $\mathcal{D}$ with a number of $V$ items (as e.g. businesses or services, movies, products, etc.) and a number of $U$ users. 
An element $e \in \mathcal{D}$ consists of a sequence of $N_e$ reviews $\*r_e = \{(\*x^e_{t},\tau^e_t, \delta^e_t, \*y^e_t)\}_{t=1}^{N_e}$, where the $t$-th review is composed of its text $\*x^e_{t}$, creation time $\tau^e_t$, inter-review time $\delta^e_t \equiv \tau^e_t-\tau^e_{t-1}$ and rating vector $\*y^e_t$. 

Such review sequences $\*r_e$ effectively define time series, and each of these can either be associate with a user $u$ (in which case we set $e=u$), or an item $v$ (in which case $e=v$).

Thus the rating vector for user $u$ is such that $\*y^u_t\in\mathbb{R}^V$, with $\*y^u_{t,v}=p$ if user $u$ rated item $v$ with rating $p$. Conversely, the rating vector for item $v$ is such that $\*y^v_t\in\mathbb{R}^U$. 
Note that both of these vectors are large and sparse. To process them efficiently we perform
dimensionality reduction via hashing, following \cite{weinberger2009feature}.

%
%
%
Our main idea is to model the user and item review sequences separately, via two independent RNNs which 
output temporal representations encoding the nonlinear relations between timing and rating of past reviews. 
We then feed these temporal representations to neural models of text, thereby yielding instantaneous review content models, while simultaneously use them to predict when are new reviews going to arrive and what are their ratings.
The model thus consists of tree interacting components: a temporal model composed of two RNNs, one for users and the other for items, which we called \textit{Dynamic Model of Review Sequences},
a neural language model which leverages the temporal representations of both user and items, and which we called \textit{Dynamic Model of Review Content}, and a \textit{Rating Model} which combines the user and item temporal representations with the review content representations to predict ratings. In what follows we dwell into the details of these building blocks. Figure \ref{fig:model} summarizes the Dynamic Review-based Recommender (DRR) model.

\subsection{Dynamic Model of Review Sequences}
\label{ssec:user_item_state}
%

Given a sequence of reviews $\*r_e$, we process each of its elements recursively via a RNN with hidden state $\*h_t^e \in \mathbb{R}^H $. At each timestep $t$, we first compute the hidden representation
\begin{equation} \label{eq:hidden-rep-transition-funct}
    \*z^e_t= \*W^{e}_{\tau} \tau^e_{t} + \*W^e_{\delta} \delta^e_t + \*W^e_{y}  \*y_t^e + \*b^e, 
\end{equation}
where $\*W^{e}_{\tau}, \*W^{e}_{\delta}, \*W^{e}_{y}$ and $\*b^{e}$ are learnable parameters and $\*z^e_t \in \mathbb{R}^E$. We then update the RNN's hidden state thus
\begin{equation}  \label{eq:transitionfunction}
    \*h^e_t = f_{\theta}^{(e)}(\*z^e_t, \*h^e_{t-1}),
\end{equation}
where $f_{\theta}^{(e)}$ is implemented by a LSTM network \cite{LSTM}.
%
%

%
Note that the superindex $e$ is used here to emphasize that we have two sets of functions namely, one for the user ($e=u$) and one for the item ($e=v$) reviews. 
%
%
The temporal representation $\*h_t^e$ thus defined not only encodes the history of ratings, but also the time lag between past reviews, thereby yielding a continuous-time representation of the dynamics.

To enforce encoding quality, we first use $\*h_t^e$ to predict the arrival time of new reviews via a simple Review Creation Model, which we shall now introduce. 
Later we will explicitly use $\*h_t^u$ and $\*h_t^v$ to predict ratings through a Rating Model.


\subsubsection{Review Creation Model} 
The inter-review times $\delta^e_{t}$ can be modeled as following an exponential distribution whose rate parameter $\lambda^{(e)}_{\theta}(\*h^e_t)$ is a function of the temporal representation $\*h_t^e$ \cite{cvejoski2019recurrent, cvejoski2020recurrent}. 
In practice we approximate the function $\lambda_{\theta}^{(e)}: \mathbb{R}^H \rightarrow \mathbb{R}_{>0}$ with a multi-layer perceptron. 
The log-likelihood of the Review Creation Model is then
\begin{equation}    \label{eq:review_creation_loss}
\begin{split}
    \log{p(\mathbf{\delta}^e)}=&\sum^{N_e}_{t=1}\log{p_{\theta}(\delta_{t+1}^e|\*h_t^e)}\\
    =&\sum^{N_e}_{t=1}\left(\log\lambda_\theta^{(e)}\left(\*h_t^e\right) -\lambda_\theta^{(e)}\left(\*h_t^e\right)\delta_{t+1}^e\right).
\end{split}
\end{equation}
Note that predicting the arrival times of new reviews can be done by either sampling the exponential distribution, or using the mean of the distribution directly. In our experiments we use the mean of the distribution. 
\subsection{Dynamic Model of Review Content} 
\label{ssec:BoW}
%
%
Consider the $t$-th review in the sequence $\*r_e$, whose text content is given by $\*x^e_t = (\*w^{e,t}_0, \*w^{e,t}_1, \dots, \*w^{e,t}_{L^e_t})$, 
where $\*w^{e,t}_j$ and $L^e_t$ label the $j$-th word and the number of words in that review, respectively. 
To capture how the review content changes within $\*r_e$, we define the probability of observing the word sequence $\*x^e_t$ at the $t$-th review as the conditional probability $p(\*x^e_t | \*h_{t-1})$.
Here we define the global temporal representation $\*h_t$ encoding the nonlinear relations between timing and ratings of past reviews as $\*h_t \equiv \mbox{concat}([\*h_t^u, \*h_t^v])$,
with $\*h_t^{u, v}$ defined in Eq. \ref{eq:transitionfunction}. 

Note that when processing the dataset $\mathcal{D}$, the modeling of review content does not need to differentiate between user and item. We therefore drop the superindex $e$ in what follows.

Below we present two models for $p(\*x_t | \*h_{t-1})$, one based on a Bag-of-Words (BoW) representation, and another on an autoregressive language model. 
Both models will be trained by maximising $\log p(\*x_t | \*h_{t-1})$. 
These language models will ultimately allow us to define a vector representation $\bar{\*s}_t$, summarizing the content of the $t$-th review, which we will later use as input to our Rating Model.

\subsubsection{Bag-of-Words Neural Review Model} 

We assume the words in $\*x_t$ are generated independently, conditioned on $\*h_{t-1}$,
that is $p(\*x_t | \*h_{t-1}) = \prod_j^{L_t} p_{\theta}(\*w^{t}_j | \*h_{t-1})$,
%
where we follow \cite{miao2016neural} and  write the probability over words as
\begin{equation}     \label{eq:word_prob}
    \begin{split}
        p_\theta(\*w^t_j|\*h_{t-1})&=\frac{\exp{\{-a(\*w^t_j,\*h_{t-1})\}}}{\sum_{k=1}^V \exp{\{-a(\*w^t_k,\*h_{t-1})\}}},   \\
        a(\*w^t_j,\*h_{t-1}) &= -\*h^{\top}_{t-1} \*R \, \*w^t_j - \*b \, \*w^t_j, 
\end{split}
\end{equation}
with $\*R \in \mathbb{R}^{2H \times V}$ and $\*b \in \mathbb{R}^V$ trainable parameters, $\*h_t = \mbox{concat}([\*h_t^u, \*h_t^v])$ and $\*w^t_j$ the one-hot representation of the $j$-th word in $\*x_t$.


We define the summary representation for $\*x_t$ as the Bag-of-Words (BoW) representation $\bar{\*s}_t \in \mathbb{R}^V$, where $V$ is the vocabulary size \cite{NIPS2009_3856}. 

\subsubsection{Autoregressive Review Model} In contrast to the BoW model above, autoregressive language models approximate the probability over the word sequence $\*x_t$ as \cite{Mikolov}
%
%
\begin{equation}
p(\*x_t|\*h_{t-1}) = \prod^{L_{t}}_{j=1} \, p_{\theta}(\mathbf{w}^{t}_j|\mathbf{w}^{t}_{<j}, \*h_{t-1}),
\label{eq:LM}
\end{equation}
where $\*w^t_{<j}$ labels all words previous to $\*w^t_{j}$.

The conditional probability above depends on both $\*w^t_{<j}$ and the global temporal representation $\*h_t$.
To model it we take an approach akin to that of the variational autoencoders of text \cite{bowman2016generating}.
That is, we first concatenate $\*h_t$ with all word embeddings in $\*x_t$, 
i.e. we define $\tilde{\*w_j} = \mbox{concat}[\*w_j, \*h_{t-1}]$, and then process the new vector sequence with a RNN with hidden state $\*s_k^t \in \mathbb{R}^S$, whose update equation reads
$\*s^t_j = g_{\theta}(\tilde{\*w_j}, \mathbf{s}^{t}_{j-1})$.
Here $g_{\theta}$ is implemented by a LSTM network, with equations similar to those below \Eqref{eq:transitionfunction}.

The distribution $p_{\theta}$ is then defined as a categorical distribution over a vocabulary of size $V$, whose class probabilities 
%
%
are given by
    $\boldsymbol{\pi}^t_j = \mbox{softmax}(\*W \, \*s^t_j)$,
where $\*W \in \mathbb{R}^{V \times S}$ is a learnable matrix.
%
%
%


We now define the summary representation for $\*x_t$ as a weighted sum over word representations 
%
%
    $\bar{\*s}_t = \sum_j^{L_t}\alpha_j^{t} \*s^{t}_j$
%
where the $j$-th weight $\alpha_j^{t}$ is calculated with the \textit{gated attention mechanism} proposed in \cite{ilse2018attention}
\begin{equation} \label{eq:gated_attention}
    \begin{split}
    \alpha_j^{t}=&\mbox{softmax}(\*k_j^{\top} \*q), \\
    \*k_j=&\tanh(\*M_1 \, \*s^{t}_j + \*b_1)\odot\sigma(\*M_2 \, \*s^{t}_j+\*b_2),
    \end{split}
\end{equation}
where $\*M_1,\*M_2\in\mathbb{R}^{A\times S}$ and $\*b_1,\*b_2\in\mathbb{R}^{A}$ are learnable parameters, $\*q\in\mathbb{R}^{A}$ can be interpreted as a learnable global query, $\odot$ denotes element-wise multiplication and $\sigma(\cdot)$ denotes the sigmoid function. 
This type of attention is introduced to solve the problem of the limited expressiveness of the $\tanh(\cdot)$ to capture complex relations, due to the fact of approximate linearity in the region $[-1,1]$. 

As we shall see below, this attentive summary representation allows us to track the most relevant words affecting the rating of a given item as time evolves. 
\subsection{Combining temporal and summary representations}

Given the temporal representations for user and item reviews (i.e. $\*h_t^u$, $\*h_t^v$), and the summary representation for review content $\bar{\*s}_t$,
we want to predict the rating $\hat{y}^{uv}_t \in \mathbb{R}$ that user $u$ gives to item $v$. 
There is, however, still the question of how to combine $\*h_t^u$ and $\*h_t^v$ with $\bar{\*s}_t$. After exploring different possibilities we found two optimal solutions namely,

\subsubsection{DRR-BoW}
For the Bow Neural Review Model we augment \Eqref{eq:hidden-rep-transition-funct} and define 
    $\tilde{\*z}_t^e = \*z_t^e + \*W^e_s \, \bar{\*s}_t$,
where $\*W^e_s \in\mathbb{R}^{H \times S}$ is an additional learnable weight, to get
$\tilde{\*h}_t^e = f_{\theta}^{(e)}(\tilde{\*z}^e_t, \tilde{\*h}^e_{t-1})$,    
where $f_{\theta}^{(e)}$ remains the same as in \Eqref{eq:transitionfunction}. The new representation $\tilde{\*h}_t^e$ now encodes the nonlinear interaction between  timing, rating \textit{and text} of past reviews.

\subsubsection{DRR-LM}
For the Autoregressive Review Model we instead define
\begin{equation} \label{eq:DRR-LM-C}
    \tilde{\*h}_t^e = \*W^{(e)} \mbox{concat}([\*h_t^e, \bar{\*s}_t]) + \*b^{(e)},
\end{equation}
 with $\*W^{(e)} \in\mathbb{R}^{H \times (H+S)}, \*b^{(e)}\in\mathbb{R}^H$ learnable. %
 The resulting representation $\tilde{\*h}_t^e$ also encodes the interaction between  timing, rating and text, albeit through a different route.

\subsection{Rating Model}
\label{ssec:Rating Model}

We have now all ingredient to predict the rating $\hat{y}^{uv}_t \in \mathbb{R}$ that user $u$ gives to item $v$. We compute $\hat{y}^{uv}_t$ with a factorization machine (FM) \cite{rendle2010factorization}, here defined as
\begin{equation}    \label{eq:fm}
    \begin{split}
    \hat{y}^{uv}_{t+1}(\*h) &= w_0 + \sum_{i=1}^{2H} w_i h_i + \sum_{i=1}^{2H}\sum_{j=i+1}^{2H}\langle\*v_i,\*v_j\rangle h_i h_j, \\
    \end{split}
\end{equation}
where $\*h\in\mathbb{R}^{2H} \equiv  \mbox{concat}([\tilde{\*h}_t^u, \tilde{\*h}_t^v])$, $w_0\in\mathbb{R}$, $\*w\in\mathbb{R}^{2H}$ and $\*V\in\mathbb{R}^{2H \times K}$ are learnable parameters, $K$ is set to 10 and $\langle\cdot,\cdot\rangle$ denotes dot product. 

We choose the loss function of the Rating Model to be the mean square error function between $\hat{y}^{uv}_t$ and our prediction $\hat{y}^{uv}_{t}(\*h)$.

\subsection{DRR Loss function}
%
The complete loss function of the DRR model has therefore three components: the loss of the Rating Model, the loss of the Dynamic Model of Review Sequences, which is the negative log-likelihood of an exponential, and the loss of the Dynamical Model of Review Content, which is the negative log-likelihood of our word sequence model.
Explicitly we write
\begin{equation}     \label{eq:loss}
\begin{split}
        \mathcal{L} =& \frac{1}{|\mathcal{D}|}\sum_{u,v\in\mathcal{D}} \sum_t (y^{uv}_{t}-\hat{y}^{uv}_{t=1})^2 \\ -& \lambda_1 \sum_{e\in \mathcal{D}}\sum^{N_e}_{t=1}\log{p_{\theta}(\delta_{t+1}^e|\*h_t^e)} 
        - \lambda_2 \sum_t \log p_{\theta}(\*x_t | \*h_{t-1}),
\end{split}
\end{equation}
where $\lambda_1, \lambda_2 \in \mathbb{R^+}$ are hyperparameters.
\begin{table*}[t]
\caption{Datasets statistics. The mean and the standard deviation of the number of reviews, sentences and words per review with respect to the user and item.}
\centering
\begin{tabular}{lcrcrcrcr}
            \cline{2-9}\cline{2-9}\cline{2-9}\cline{2-9}
          & \multicolumn{2}{c}{\textbf{Automotive}}                           & \multicolumn{2}{c}{\textbf{Digital Music}}                       & \multicolumn{2}{c}{\textbf{Tools and Home}}                      & \multicolumn{2}{c}{\textbf{Pet Supplies}}                        \\ \cline{2-9}
          & \multicolumn{2}{c}{\textbf{user/item}}                            & \multicolumn{2}{c}{\textbf{user/item}}                           & \multicolumn{2}{c}{\textbf{user/item}}                           & \multicolumn{2}{c}{\textbf{user/item}}                           \\
          & \textbf{mean}                  & \multicolumn{1}{c}{\textbf{std}} & \textbf{mean}                 & \multicolumn{1}{c}{\textbf{std}} & \textbf{mean}                 & \multicolumn{1}{c}{\textbf{std}} & \textbf{mean}                 & \multicolumn{1}{c}{\textbf{std}} \\
            \specialrule{.1em}{.05em}{.05em} 
reviews   & \multicolumn{1}{r}{6.1/9.3}    & 1.7/5.5                          & \multicolumn{1}{r}{7.8/9.0}   & 6.2/6.5                          & \multicolumn{1}{r}{7.8/10.6}  & 5.3/9.0                          & \multicolumn{1}{r}{7.44/13.7} & 4.4/14.2                         \\
sentences & \multicolumn{1}{r}{8.7/9.7}    & 6.3/7.5                          & \multicolumn{1}{r}{6.3/4.9}   & 11.9/10.2                        & \multicolumn{1}{r}{8.4/7.5}   & 8.1/7.8                          & \multicolumn{1}{r}{7.8/6.5}   & 7.7/6.9                          \\
words     & \multicolumn{1}{r}{89.8/101.6} & 68.3/82.5                        & \multicolumn{1}{r}{52.5/38.2} & 106.4/94.6                       & \multicolumn{1}{r}{80.3/70.5} & 85.6/83.3                        & \multicolumn{1}{r}{68.9/55.9} & 71.1/66.2    \\
    \hline
\end{tabular}

\label{tab:arrivals_stats}
\end{table*}

\section{Causality}
\label{ssec: causal_non-causal}
%
%
By construction, both DRR-BoW and DRR-LM models above preserve causality --- the models do not use \textit{any} information from the future to predict ratings. As mentioned in the introduction, however, most recommender system models that leverage review content use the review $\*x^v_{t+1}$, written by user $u$, to predict the rating $y^{uv}_{t+1}$ given by this same user to the item $v$. 
In order to fairly compare our methodology with such models, we use the degrees of freedom available within the definition of the DRR-LM model and redefine 
\begin{equation} \label{eq:DRR-LM-NC}
    \tilde{\*h}_t^e = \*W^{(e)} \mbox{concat}([\*h_t^e, \bar{\*s}_{t+1}]) + \*b^{(e)}.
\end{equation}

This new representations encodes $\bar{\*s}_{t+1}$, the summary representation of the review whose rating it predicts, and breaks causality.
Below we refer to the model using the causal representation \Eqref{eq:DRR-LM-C} as DRR-LM-C, whereas we denote the model using the non-causal expression \Eqref{eq:DRR-LM-NC} as DRR-LM-NC.
%
%

Naturally, the causal model is to be preferred as we normally do not have review content about the item whose rating we want to predict. Nevertheless, we shall see that the non-causal model lends itself when one is interested in tracking the words which most affect the rating of a given item as time evolves.

\section{Experiments and Results}
\label{sec:ER}
\begin{table*}[!t]
  \caption{Mean-square error on the rating prediction (* results taken from \cite{Dong2019AsymmetricalHN}).}
\centering
\begin{tabular}{c|c|cccc|ccc}
\toprule
& \multicolumn{1}{c}{\textbf{static}}
& \multicolumn{4}{c}{\textbf{non-causal models}}                
& \multicolumn{3}{c}{\textbf{causal-models}}          \\
\multicolumn{1}{c}{\textbf{Datasets}} & \textbf{PMF*} & \textbf{DeepCoNN*} & \textbf{D-ATT*} & \textbf{AHN*}   & \textbf{DRR-LM-NC} & \textbf{RRN} & \textbf{DRR-BoW} & \textbf{DRR-LM-C} \\
\midrule
\textbf{A}                   & 0.9187        & 0.7809             & 0.7654    & \textbf{0.7314} & 0.7791             & 1.0927       & \textbf{0.7838}           & 0.8171            \\
\textbf{DM}                & 0.8788        & 0.8754             & 0.8506          & 0.8172          & \textbf{0.7250}    & 0.7961       & \textbf{0.7723}     & 0.7801            \\
\textbf{TH}               & 1.1182        & 0.9856             & 0.9850          & 0.9671    & \textbf{0.9264}    & 1.0896       & \textbf{1.0406}           & 1.0656            \\
\textbf{PS}                & 1.4340        & 1.2598             & 1.2730          & 1.2515          & \textbf{1.0500}    & 1.1970       & \textbf{1.1734}     & 1.1918            \\ 
\hline
\end{tabular}
  \label{tab:results}
\end{table*}

\textbf{Data set}
To test our model we choose the Amazon dataset \cite{he2016ups}. We pick four 5-core subcategory datasets namely, \textit{Automotive (A)}, \textit{Digital Music (DM)}, \textit{Tools and Home (TH)} and \textit{Pet Supplies (PS)}. The review creation time is defined as the difference in days between the original timestamp and the timestamp of the first review in the dataset. Next we group reviews by day, since the granularity of the timestamps is day based. All users or items with less than 5 days (i.e. time series with less than 5 points) are removed from the dataset. The autoregressive language models use the review raw text, changed into lower case. 
Preprocessing scripts can be found at \cite{source_code}. Statistics of the preprocessed data is summarized in Table \ref{tab:arrivals_stats}.

\textbf{Training} Our model predicts ratings through the user and item dynamic representations, which come from two independent RNNs. Simply applying backpropagation through both sequences is computationally forbidden. In order to overcome this problem, we train the user and item RNNs alternately. We first freeze the parameters of e.g. the items' RNN, and only update those of the users' RNN, while back-propagating the gradients of all ratings for a user batch. The items' dynamic representations are taken to be fixed. We then repeat these operations but now with the user parameters and user representations frozen. 

\textbf{Model Configuration} We split each dataset along the time dimension into three parts: training set (80\%), validation set (10\%) and test set (10\%). We use grid search on the validation set for hyperparameter tuning. We set the hidden dimension $H$ of the temporal representation $\*h_t^e$ to 32, and the embedding dimension $E$ of $\*z^e_t$ to 100.
Regarding the review content models, we set the vocabulary size $V$ to 2000 for DRR-BoW and to 5000 for DRR-LM. 
In the latter case we also use GloVe word embeddings \cite{pennington2014glove} (these corresponds to the $\*w_j^t$ in \Eqref{eq:LM}) with dimension $300$.
For DRR-LM we also set the attention dimension $A$ to 64 and the embedding dimension $H'$ of the (concatenation of the) temporal and summary representations to $64$. 
We use Adam \cite{ADAM} with learning rate 0.0002 and $\beta_1=0.9$
and limit the review length to 150 tokens. All methods are implemented using PyTorch v1.3\footnote{https://pytorch.org/}. Source code for all models can be found at \cite{source_code}.

\begin{figure*}[h]
     \centering
     \includegraphics[width=0.30\linewidth]{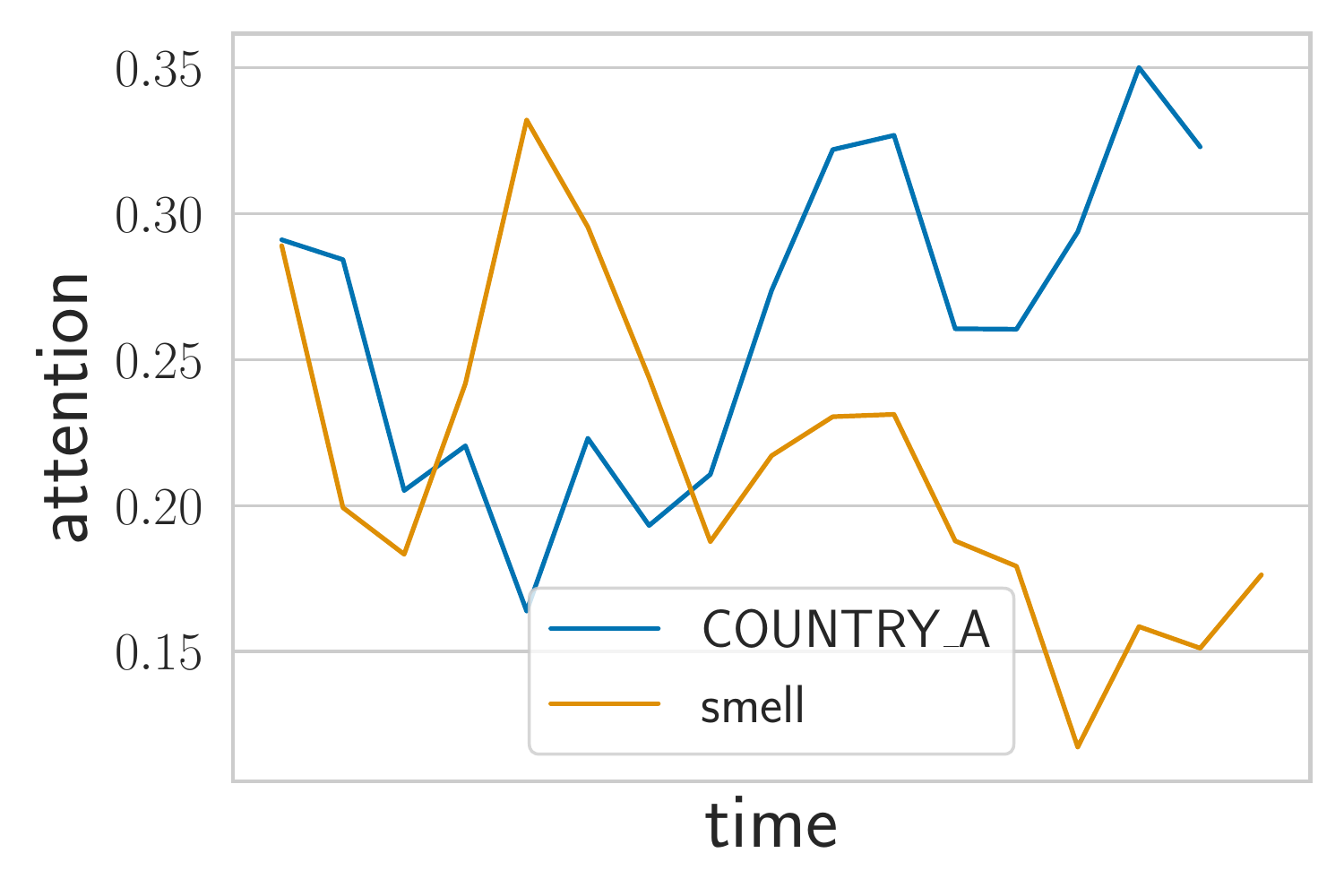}
     \hfill
     \includegraphics[width=0.28\linewidth]{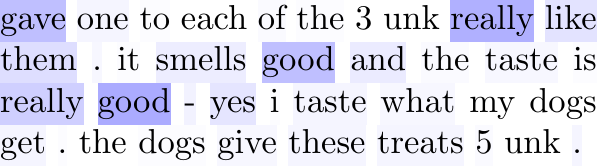}
     \hfill
     \includegraphics[width=0.30\linewidth]{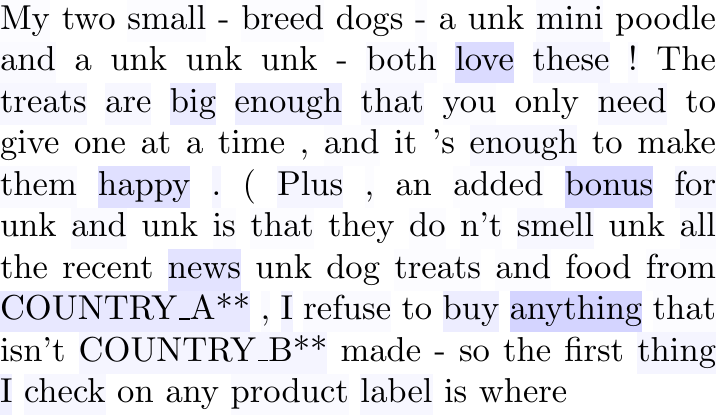}
     \includegraphics[width=0.30\linewidth]{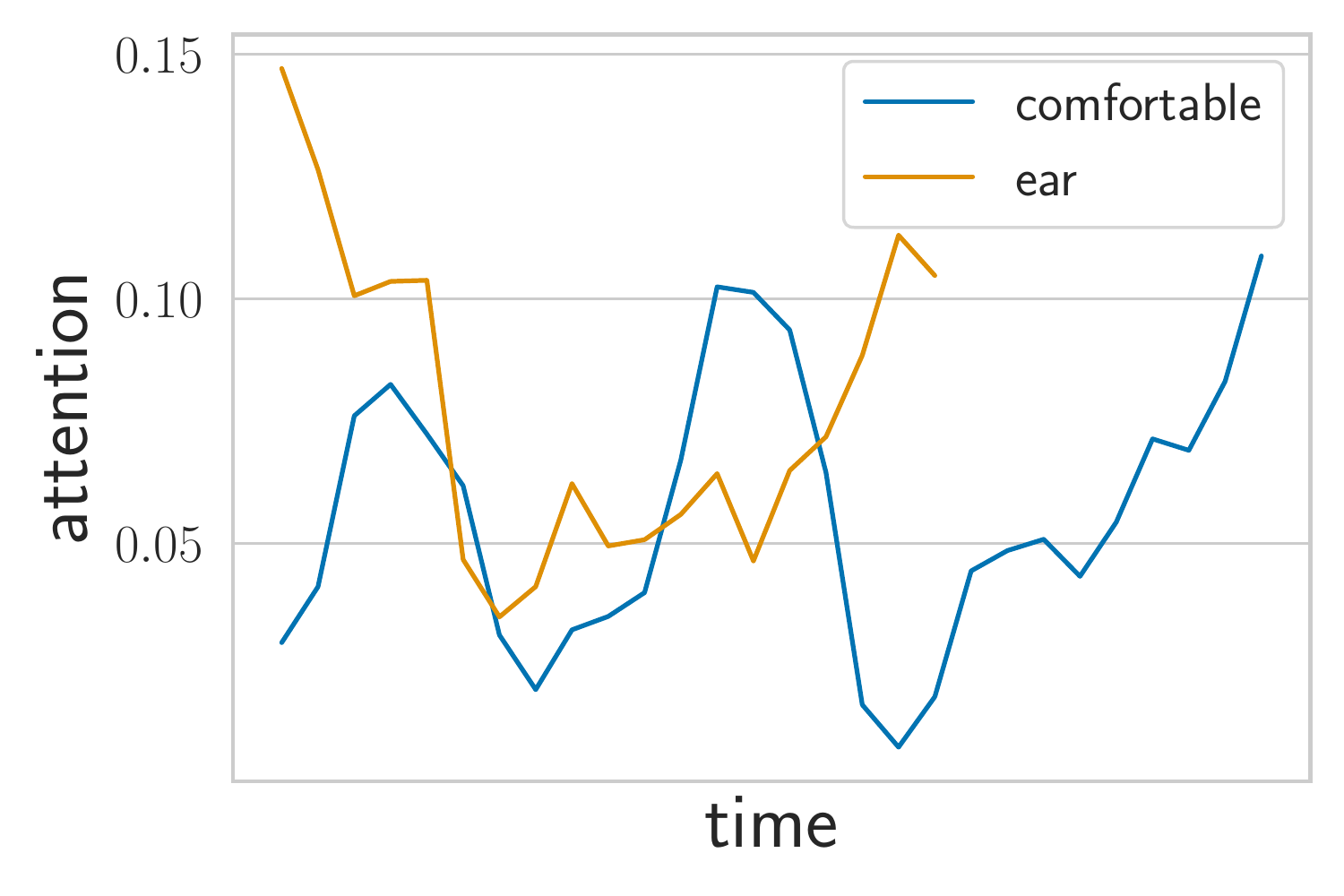}
     \hfill
     \includegraphics[width=0.30\linewidth]{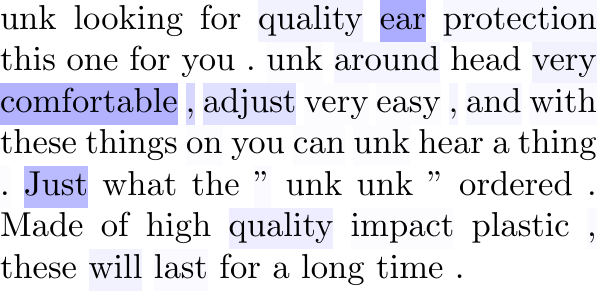}
     \hfill
     \includegraphics[width=0.30\linewidth]{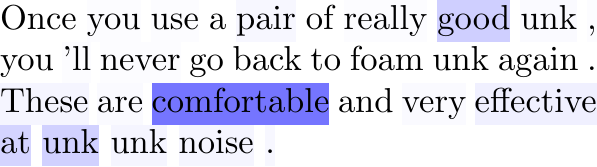}
  \caption{\textit{Upper Left:} Dynamic attention on the words 'COUNTRY\_A' and 'smell' for an item in 'Pet Supplies' dataset. \textit{Upper Middle:} Review sample from the beginning of the time series. \textit{Upper Right:} Review sample from the end of the time series. **The real names of the countries are replaced with masks 'COUNTRY\_A' and 'COUNTRY\_B' for fairness. \textit{Lower Left:} Dynamic attention on the words 'comfortable' and 'ear' for an item in the 'Tools and Home' dataset. \textit{Lower Middle:} Review sample from the beginning of the time series. \textit{Lower Right:} Review sample from the end of the time series. The darker the highlight color for a word, the higher its attention value.}
  \label{fig:att_in_time}
\end{figure*}

\textbf{Results}
Given an user and item of interest, the DRR model predicts the arrival time, rating and the probability over the word sequence of the next review,
and we optimize the model to give the best performance on the rating prediction task.
%
%

Our methodology incorporates modeling the dynamic aspects of user-item interaction with neural models of review content. To test the importance of each of these components for the problem of rating prediction, we test our models against (i) the Probabilistic Matrix Factorization (PMF) \cite{mnih2008probabilistic}, which is a static recommender system which does not model review content; (ii) the RRN \cite{RecurrentRecommender}, a causal model which learns dynamic user/item representations (albeit non-continous), but does not model review content; and (iii) three static models which do leverage review content, namely DeepCoNN \cite{zheng2017joint}, D-ATT \cite{seo2017interpretable} and AHN \cite{Dong2019AsymmetricalHN}. These last three models are non-casual since they either use the review of the item whose rate they predict, or use item reviews that have not been received by the time the item of interest was rated.
%
%
Table \ref{tab:results} shows results for all models on the chosen datasets. We use boldface to highlight best results in both causal and non-causal cases.

Let us start by focusing on the causal models. First we note that both DRR-BoW and DRR-LM-C outperform the RRN model, which confirms the known fact that review content helps in rating prediction tasks. 
We remark however that in this case the models in questions are dynamic, and it is the content of \textit{past reviews} what is successfully being used.
Interestingly, DRR-BoW beats DRR-LM-C which may hint at the fact that it is enough to know that certain key words are present in the review, as opposite to e.g. word order, to better predict the rating.
Regarding the non-causal models, DRR-LM-NC outperforms all other models in almost all datasets, which shows that one indeed needs to not only account for review content, but also for its dynamic character.
Remarkably, both causal models DRR-BoW and DRR-LM-C perform better than all their  non-causal competitors in two of the datasets (see the Digital Music and Pet Suplies rows in the table), and comparable to them in the others.

We can conclude that our models successfully learn both temporal user/item representations and review content representation which \textit{together} are useful for rating prediction.

Let us now consider the dynamic attention mechanism of the DRR-LM-NC, which allows us to e.g. follow in time the weights $\alpha_j^t$ (defined in Eq. \ref{eq:gated_attention}) of the words in the reviews for the item whose rate we aim at predicting. The higher the weight of a word, the stronger its relevance to the rating prediction.
Figure \ref{fig:att_in_time} \textit{Upper Left} shows the attention weights on the words `COUNTRY\_A' and `smell' as time evolves for a given product in the `Pet Supplies' dataset.
One can see that although at the start of the time series the word `smell' was important for determining the rating, its relevance decreases as the weight on the word `COUNTRY\_A' increases.
After a closer look at the reviews we learn that at the start of the time series most reviews were related to the smell of the product (e.g. whether the dogs were liking the product's smell). Later on, however, the manufacturing company moved the product production to COUNTRY\_A, and this event was successfully captured by our attention model.
%
%
Figure \ref{fig:att_in_time} \textit{Upper Middle} shows an example review for the item in question, from the start of the time series. 
Words with darker highlights mean here words with higher attention weight.
One can see that the word `smell' is highlighted as important. In contrast, Figure \ref{fig:att_in_time} \textit{Upper Right} displays a review sampled from the end of the time series, in which one sees the word `COUNTRY\_A' has more relevance than the word `smell'. 
Similarly, the \textit{Lower} row of Figure \ref{fig:att_in_time}  shows  the  attention  weights on the words `comfortable' and `ear' as time evolves for a given product in the ‘Tools and Home’ dataset. 
%

%
%

\section{Conclusion and Feature Work}
In this work we proposed a recommender system model which accounts for the dynamic aspects of user preferences, as reflected in their history of reviews and ratings. 
We explicitly learn continuous-time representations for both users and items, and use these to define dynamic language models for review content.
The latter provided us with review content representations which, when combined with the temporal user/item representations, proved to be useful in predicting the hidden interest of users in unknown items. 
Indeed, our results outperformed several state-of-the-art recommender system models in rating prediction tasks, in different datasets. 
%
%
%

We also introduced a new dynamic attention mechanism which allowed us to track the most relevant words for a given rating of an item of interest at a given instant of time. 

Future directions of work include developing attention mechanisms between 
reviews with different timestamps, and learning more generic dynamic representations able to characterize hidden dynamics global to all users.

\section*{Acknowledgment}

The authors of this work were supported by the Competence Center for Machine Learning Rhine Ruhr (ML2R) which is funded by the Federal Ministry of Education and Research of Germany (grant no. 01|S18038A). Part of the work was also funded by the BIFOLD-Berlin Institute for the Foundations of Learning and Data (ref. 01IS18025A and ref. 01IS18037A). We gratefully acknowledge this support.

\bibliographystyle{IEEEtran}
\bibliography{main}

\end{document}